\documentclass[9pt,conference]{IEEEtran}

\usepackage[preprint]{waspaa25}

\usepackage{bm} 
\usepackage{dirtytalk}
\usepackage{empheq}     
\usepackage{nicefrac}
\usepackage{nicematrix}
\usepackage{optidef}
\usepackage{siunitx}
\usepackage{soul}
\usepackage{stfloats}
\usepackage[subrefformat=parens,labelformat=parens,caption=false]{subfig}
\usepackage{tcolorbox} 
\usepackage{textcomp}
\usepackage{upgreek}
\usepackage{lipsum}
\usepackage{nicefrac}

\title{Soft-Constrained Spatially Selective Active Noise Control \\for Open-fitting Hearables}

\name{Tong Xiao$^{1}$, \thanks{This research was funded by the Deutsche Forschungsgemeinschaft (DFG, German Research Foundation) -- Project-ID 352015383 -- SFB 1330 C1.}
      Reinhild Roden$^{2}$, 
      Matthias Blau$^{2}$, 
      Simon Doclo$^{1}$}
\address{$^1$ Department of Medical Physics and Acoustics and Cluster of Excellence Hearing4all, \\ Carl von Ossietzky Universit\"{a}t Oldenburg, Germany \\ tong.xiao@uni-oldenburg.de, simon.doclo@uni-oldenburg.de\\
         $^2$ Institut f\"{u}r H\"{o}rtechnik und Audiologie and Cluster of Excellence Hearing4all, Jade-Hochschule, Oldenburg, Germany\\ reinhild.roden@jade-hs.de, matthias.blau@jade-hs.de\\
}

\begin{document}

\maketitle

\begin{abstract}
Recent advances in spatially selective active noise control (SSANC) using multiple microphones have enabled hearables to suppress undesired noise while preserving desired speech from a specific direction. Aiming to achieve minimal speech distortion, a hard constraint has been used in previous work in the optimization problem to compute the control filter. In this work, we propose a soft-constrained SSANC system that uses a frequency-independent parameter to trade off between speech distortion and noise reduction. We derive both time‑ and frequency‑domain formulations, and show that conventional active noise control and hard-constrained SSANC represent two limiting cases of the proposed design. We evaluate the system through simulations using a pair of open‑fitting hearables in an anechoic environment with one speech source and two noise sources. The simulation results validate the theoretical derivations and demonstrate that for a broad range of the trade-off parameter, the signal-to-noise ratio and the speech quality and intelligibility in terms of PESQ and ESTOI can be substantially improved compared to the hard-constrained design.
\end{abstract}

\section{Introduction}
\label{sec:intro}

Active noise control (ANC) hearables aim to create a quiet environment by using secondary sources to generate anti-noise, minimizing sound when superimposed on the primary noise (also referred to as leakage)\cite{Elliott2000, Hansen2012active}. Based on their fit, hearables can be categorized as closed-fitting (completely occluding the ear), open-fitting (partially occluding the ear), and open-ear (no occlusion; e.g., smart glasses). While open-fitting and open-ear designs reduce the occlusion effect and improve physical comfort, they also lead to increased leakage. Recent research focuses on intelligent ANC hearables with spatial selectivity for complex acoustic environments involving multiple sound sources from different directions\cite{kajikawa2012recent, Chang2016Listening, gupta2022augmented, Serizel2010integrated, Patel2020design, xiao2023spatial}. In such scenarios, listeners often want to focus on sound from a particular direction (e.g., speech from the front) while suppressing noise leakage from other directions.

Modern ANC hearables are commonly equipped with multiple microphones, i.e., microphones on the exterior of the hearable and error microphones in the interior close to the eardrum. On the one hand, conventional ANC algorithms suppress all leakage measured by the inner error microphones, including desired speech\cite{Elliott2000, Hansen2012active,benois2022optimization}. On the other hand, multi-microphone noise reduction algorithms, e.g., minimum variance distortionless response (MVDR) beamforming~\cite{vanveen1988, Doclo2015, gannot2017consolidated}, perform spatial filtering by reducing all noise while preserving speech from the desired direction. However, these algorithms typically ignore the leakage and do not exploit the inner error microphones. Recent work on spatially selective active noise control (SSANC) integrates spatial filtering into ANC systems, thus preserving speech from a desired direction while actively suppressing sound from undesired directions~\cite{Serizel2010integrated, Serizel2011output, SERIZEL2013speech, Dalga2011combined, Dalga2012theoretical, Ho2018integrated, Patel2020design, xiao2023spatial, Zhang2023time, Zhang2025spherical, Xiao2024icassp, Xiao2025fa}. To achieve minimal speech distortion, the SSANC algorithm in~\cite{xiao2023spatial} imposes a hard constraint in the optimization problem for the control filter. For this algorithm, it has been shown in~\cite{Xiao2024icassp} that when aiming to preserve the speech component in an outer reference microphone signal, a delay equal to the acoustic propagation delay between the reference microphone and the inner error microphone for the desired source is necessary to satisfy causality. Furthermore, it has been shown in~\cite{Xiao2025fa} that using acausal relative impulse responses in the filter optimization is beneficial.

In this paper, we introduce a soft-constrained SSANC system that employs a single frequency-independent parameter to adjust the trade‑off between speech distortion and noise reduction. Although soft-constrained design procedures have been explored for multi-microphone speech enhancement~\cite{doclo2002gsvd, Spriet2004spatially, nordholm2005adaptive, Doclo2007frequency, braun2015residual}, e.g., speech distortion weighted multichannel Wiener filter (SDW-MWF), and for feedback suppression~\cite{Schepker2020acoustic}, they have rarely been studied for ANC. We derive both time‑ and frequency‑domain formulations for the conventional ANC, the hard-constrained SSANC, and the proposed soft-constrained SSANC. For the frequency-domain formulations, we show that conventional ANC and hard‑constrained SSANC correspond to limiting cases of the soft‑constrained design, validating these theoretical relationships through causal time-domain simulations. For the considered setup, simulation results demonstrate that for a broad range of the trade-off parameter, the proposed soft-constrained SSANC outperforms both hard-constrained SSANC and conventional ANC.

\section{Signal Model}
\label{sec:time_domain_signal_model}

\begin{figure}[t]
    \centering
    \includegraphics[width=0.88\linewidth]{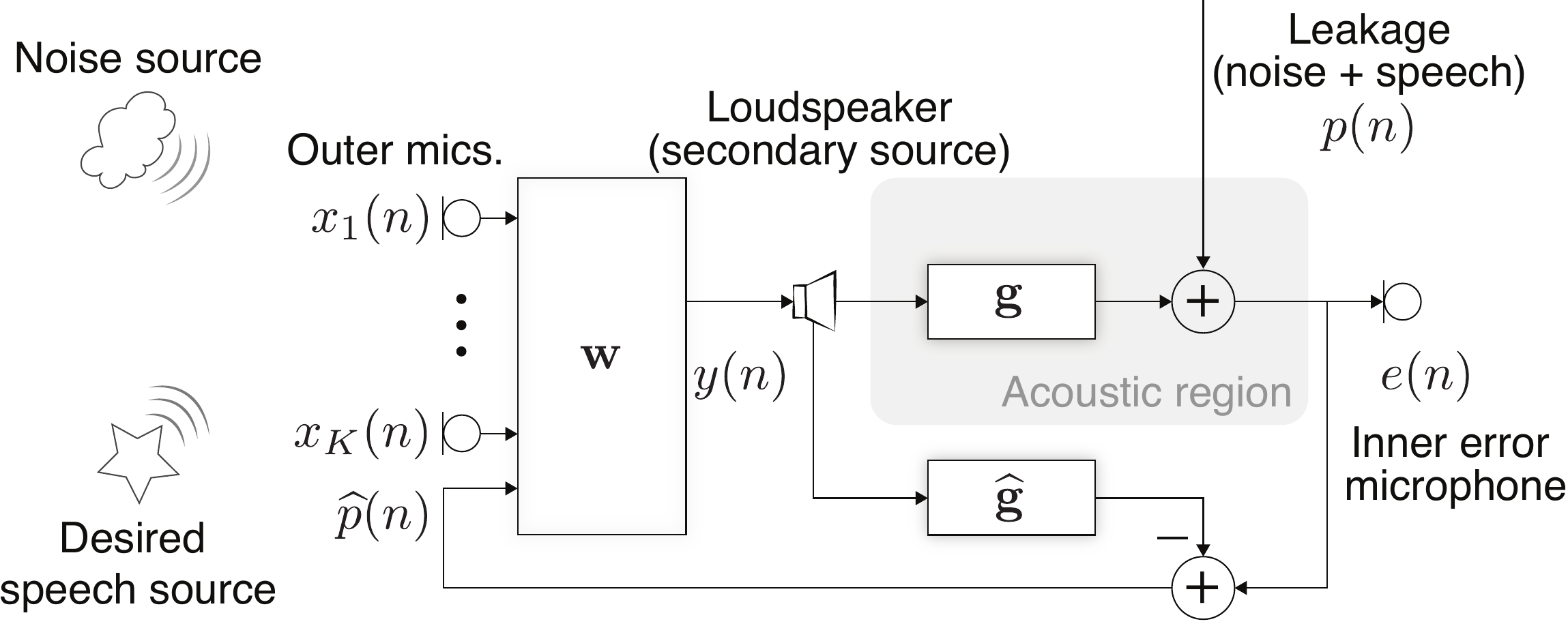}
    \caption{Block diagram of an ANC system with $K$ outer microphones, one inner error microphone and one loudspeaker (i.e., secondary source). The control filter is denoted by $\mathbf{w}$, the secondary path is denoted by $\mathbf{g}$, and its estimate by $\widehat{\mathbf{g}}$.}
    \label{fig:ssanc}
    \vspace{0pt}
\end{figure}

As shown in~\cref{fig:ssanc}, we consider a hearable with $K$ outer microphones. Without loss of generality, we consider one loudspeaker as the secondary source and one inner error microphone, resulting in a total of $K+1$ microphones. 
We assume that the acoustic feedback paths between the loudspeaker and the outer microphones are known, such that acoustic feedback can be canceled. In addition, we assume that an estimate of the secondary path between the loudspeaker and the inner error microphone is available.

The inner error microphone signal $e(n)$, with $n$ the discrete time index, is given by
\begin{align}
        e(n) &= p(n) + ( \mathbf{G}\mathbf{w})^\mathcal{T}{\mathbf{x}}(n) , \label{eq:e_dGwx}
\end{align}
where $(\cdot )^\mathcal{T}$ denotes the transpose operator. The leakage (including noise and desired speech) at the inner error microphone is denoted by $p(n)$, and the anti-noise component at the inner error microphone is given by $( \mathbf{G}\mathbf{w})^\mathcal{T}{\mathbf{x}}(n)$, where ${\mathbf{w}}$ is the stacked control filter, ${\mathbf{x}}(n)$ is the stacked input vector, and $\mathbf{G}$ represents the secondary path convolution matrix.
The stacked control filter ${\mathbf{w}}$ is defined as
\begin{subequations}
\begin{align}
    \mathbf{w} &= [ \mathbf{w}^\mathcal{T}_1 \;\;\; \mathbf{w}^\mathcal{T}_2 \;\;\; \dots \;\;\; \mathbf{w}^\mathcal{T}_{K+1} \, ]^\mathcal{T} \in \mathbb{R}^{(K+1)L_w} ,
    \\
    \mathbf{w}_k &= \left[{w_{k,0}} \ {w_{k,1}} \ \dots \ {w_{k,{L_w}-1} } \right]^\mathcal{T} \in \mathbb{R}^{L_w},  
\end{align}
\end{subequations}
where $L_w$ denotes the control filter length for each channel. 
The convolution matrix of the secondary path is defined as
\begin{subequations}
\begin{align}
    \mathbf{G} &= \mathrm{blkdiag}\left({\mathbb{G}}  \dots  {\mathbb{G}} \right) \in \mathbb{R}^{(K+1)L \times (K+1)L_w},  \label{eq:G_tilde_multi}
\\ 
    {\mathbb{G}} &= 
        \begin{bNiceMatrix}
                g_0       &  \cdots  & 0         \\[-2pt]
                \vdots    &  \ddots  & \vdots    \\[-2pt]
                g_{L_g-1} &  \ddots  & g_0       \\[-2pt]
                \vdots    &  \ddots  & \vdots    \\[-0pt]
                0         &  \cdots  & g_{L_g-1}
        \end{bNiceMatrix}
        \in \mathbb{R}^{L \times L_w} ,
    \label{eq:G_hat}
\end{align}
\end{subequations}
where $L=L_g+L_w-1$, and $L_g$ denotes the secondary path filter length.
As input signals to the control filter we consider the $K$ outer microphone signals $\mathbf{x}_k(n)$, $k=1, \dots , K$, and an estimate of the leakage $\widehat{\mathbf{p}}(n)$, i.e., the stacked input vector ${\mathbf{x}}(n)$ is defined as
\begin{align}
    {\mathbf{x}}(n) &= [\mathbf{x}_{1}^\mathcal{T}(n) \, \dots \, \mathbf{x}_{K}^\mathcal{T}(n) \,\, \widehat{\mathbf{p}}^\mathcal{T}(n) ]^\mathcal{T} \in \mathbb{R}^{(K+1)L} , \label{eq:x_tilde_multi}
\end{align}
with
\begin{subequations}
\begin{align}
    \mathbf{x}_k(n) &= \left[ x_k(n) \ \dots \ x_k(n-L+1) \right]^\mathcal{T}, \label{eq:x_k_vec}
\\ 
    \widehat{\mathbf{p}}(n) &= \left[ \,\, \widehat{p}\,(n) \, \ \dots \ \,\, \widehat{p}\,(n-L+1) \right]^\mathcal{T}.
\end{align}
\end{subequations}
The estimated leakage $\widehat{p}(n)$ can be computed from the inner error microphone signal $e(n)$, the loudspeaker signal vector $\mathbf{y}(n)=[y(n) \dots y(n-L_g+1)]^\mathcal{T}$, and an estimate of the secondary path $\widehat{\mathbf{g}}$ as
\begin{equation}
\widehat{p}(n)=e(n)-\widehat{\mathbf{g}}^\mathcal{T} \mathbf{y}(n) .
\label{eq:d_hat}
\end{equation}
Assuming a perfect estimate of the secondary path to be available, i.e., $\widehat{\mathbf{g}} = \mathbf{g} = [ {g}_0\ {g}_1\ \allowbreak \dots\ \allowbreak {g}_{L_g-1} ]^\mathcal{T}$, the leakage can be written as $p(n) = \widehat{p}(n) = \mathbf{q}^\mathcal{T}\mathbf{x}(n)$, with 
\begin{subequations}
\begin{align}
    \mathbf{q} &= [ \, \mathbf{0}^\mathcal{T} \ \ldots \ \mathbf{0}^\mathcal{T} \ \bm{\updelta}^\mathcal{T} ]^\mathcal{T} \in \mathbb{R}^{(K+1)L} ,    \label{eq:delta_tilde_multi}
\\
    \bm{\updelta} &= \left[ \, 1 \;\;\;\; 0 \;\;\; \dots \;\;\; 0 \;\; \right]^\mathcal{T}  \in \mathbb{R}^{L} .
\end{align}
\end{subequations}
Hence, the inner error microphone signal in~\labelcref{eq:e_dGwx} can be rewritten as
\begin{align}
    \Aboxed{
        e(n) = (\mathbf{q} + \mathbf{G}\mathbf{w})^\mathcal{T}{\mathbf{x}}(n) .
        }
        \label{eq:e_qGwx} 
\end{align}

The frequency-domain representations of $\mathbf{x}(n)$ and $\mathbf{w}$ are
\begin{subequations}
    \begin{align}
        \mathbf{x}(\omega) &= [X_1(\omega) \, \dots \, X_K(\omega) \;\;\;\,\, \widehat{P}(\omega) \;\;\;\;]^\mathcal{T} \in \mathbb{C}^{K+1},
        \\
        \mathbf{w}(\omega) &= [W_1(\omega) \, \dots \, W_K(\omega) \, W_{K+1}(\omega) ]^\mathcal{T} \in \mathbb{C}^{K+1},
    \end{align}
\end{subequations}
where $\omega$ denotes the radian frequency. Similar to~\labelcref{eq:e_qGwx}, the frequency-domain inner error microphone signal is
\begin{align}
    E(\omega) 
        = \left[ \mathbf{q}^\mathcal{H}_{\, \omega} + G(\omega) \mathbf{w}^\mathcal{H}(\omega) \right] \mathbf{x}(\omega) , 
        \label{eq:ssanc_e_freq}
\end{align}
where $(\cdot)^\mathcal{H}$ denotes the Hermitian transpose, and $\mathbf{q}_{\, \omega} = [0 \dots 0 \; 1 \, ]^\mathcal{T}  \in \mathbb{R}^{K+1}$. $G(\omega)$ denotes the transfer function of the secondary path.

\section{Conventional ANC and Hard-constrained SSANC in Time Domain}
\label{sec:unconstrained_anc_and_hard_constrained_ssanc_in_time_domain}
The objective of conventional ANC is to minimize the power of the error microphone signal~\cite{Elliott2000, Hansen2012active}, i.e.,
\begin{align}
    \Aboxed{
    \min_{\mathbf{w}} \quad \mathcal{E} \left\{e^{2}(n)\right\} + \beta \| \mathbf{w} \|_2^2 ,
    }
    \label{eq:anc_optimization_time_both}
\end{align}
where $\mathcal{E} \{\cdot\}$ denotes the mathematical expectation operator. $\beta$ is a regularization term. $\| \cdot \|_2$ denotes the $L_2$-norm.
Using~\labelcref{eq:e_qGwx}, the solution of~\labelcref{eq:anc_optimization_time_both} is found to be 
\begin{align}
    \mathbf{w}_\mathrm{ANC} = - \bm{\Phi}_{\mathbf{rr}}^{-1}  \bm{\upphi} ,
\end{align}
\vspace{-2mm}
where
\begin{subequations}
    \begin{align}
        \bm{\Phi}_{{\mathbf{r}} {\mathbf{r}}} &= \mathbf{G}^\mathcal{T} \bm{\Phi}_{{\mathbf{x}} {\mathbf{x}}} \mathbf{G} + \beta \mathbf{I} ,
        \\
        \bm{\Phi}_{\mathbf{xx}} &= \mathcal{E} \{{\mathbf{x}}(n) {\mathbf{x}}^\mathcal{T}(n) \} ,
        \\
        \bm{\upphi} &= \mathbf{G}^\mathcal{T} \bm{\Phi}_{{\mathbf{x}} {\mathbf{x}}} \mathbf{q} \, ,
    \end{align}
\end{subequations}
where $\mathbf{I}$ denotes the identity matrix. It should be noted that both desired speech and noise are minimized in conventional ANC.

In~\cite{xiao2023spatial, Xiao2024icassp, Xiao2025fa}, an SSANC system has been proposed where the objective is to minimize the power of the inner error microphone signal while preserving the delayed desired speech component of an outer reference microphone signal. This can be achieved by imposing the constraint
\begin{align}
    e_s(n) =x_{\mathrm{ref},s}(n-\Delta) ,
    \label{eq:es_alpha_x_ref_time}
\end{align}
where $(\cdot)_s$ denotes the speech component of a signal and $\Delta$ represents a delay.
This constraint can be reformulated as~\cite{Xiao2025fa}
\begin{align}
    \mathbf{H} (\mathbf{q} + \mathbf{G}\mathbf{w} )  = \bm{\updelta}_{\Delta} ,
    \label{eq:HqGw_delta}
\end{align}
where
\begin{subequations}
    \begin{align}
        \mathbf{H} &= \left[\mathbf{H}_1 \ \dots \ \mathbf{H}_{K+1} \right] \in \mathbb{R}^{(L_a+L_h+L-1) \times (K+1)L}  \label{eq:H_time},
        \\[0.1em]
        \mathbf{H}_k &=  \!
            \begin{bNiceMatrix}
            h_{k,-L_a} & \!\! \cdots   \!\!  & 0               \\[-4pt]
            \vdots     & \!\! \ddots   \!\!  & \vdots          \\[-4pt]
            h_{k,L_h-1}& \!\! \ddots   \!\!  & 0               \\[-4pt]
            0          & \!\! \ddots   \!\!  & h_{k,-L_a}      \\[-4pt]
            \vdots     & \!\! \ddots   \!\!  & \vdots          \\[0pt]
            0          & \!\! \cdots   \!\!  & h_{k,L_h-1}
            \end{bNiceMatrix}
            \in \mathbb{R}^{(L_a+L_h+L-1) \times L}  ,  \label{eq:Hk_time}
        \\[0.1em]
        \bm{\updelta}_{\Delta} &= [ \, \underbrace{0 \, \dots \, 0}_{L_a} \, \underbrace{0 \, \dots \, 0}_{\Delta}  \underbrace{1 \; 0 \, \dots \, 0  }_{L_h+L-1-\Delta} ]^\mathcal{T} \in \mathbb{R}^{L_a+L_h+L-1}  ,  \label{eq:delta_Delta}
    \end{align}
\end{subequations}
$\mathbf{H}_{k}$ is the convolution matrix of the relative impulse response (ReIR) between the $k$-th microphone and the reference microphone, with $L_a$ and $L_h$ denoting the length of the anti-causal and causal parts of the ReIR, respectively~\cite{Xiao2025fa}. It should be particularly noted that although the ReIRs are acausal, the control filter $\mathbf{w}$ is still causal.

To satisfy the distortionless constraint in~\labelcref{eq:HqGw_delta}, a hard constraint is added to the optimization problem in~\labelcref{eq:anc_optimization_time_both}, i.e.,
\begin{subequations}
\label{eq:ssanc_optimization_problem_both} 
\begin{empheq}[box=\fbox]{align} 
    \min_{\mathbf{w}} \quad & \mathcal{E} \left\{e^{2}(n)\right\} + \beta \| \mathbf{w} \|_2^2 \label{eq:ssanc_optimization_problem} 
    \\
    \mathrm{subject \ to} \quad & \mathbf{H} (\mathbf{q} + \mathbf{G}\mathbf{w} ) = \bm{\updelta}_{\Delta} . \label{eq:ssanc_optimization_problem_constraint}
\end{empheq}
\end{subequations}
The solution is found to be~\cite{xiao2023spatial, Xiao2024icassp, Xiao2025fa}
\begin{align}
        \mathbf{w}_\mathrm{hard} =   
        - \bm{\Phi}_{\mathbf{rr}}^{-1}  \bm{\upphi} 
        + \mathbf{A} \left(\bm{\updelta}_{\Delta} - \mathbf{H} \mathbf{q} + \mathbf{H} \mathbf{G} \bm{\Phi}_{\mathbf{rr}}^{-1} \bm{\upphi} \right)  ,
    \label{eq:w_ssanc_time_hard}
\end{align}
with 
\begin{align}
    \mathbf{A} = 
        \bm{\Phi}_{\mathbf{rr}}^{-1} \mathbf{G}^\mathcal{T} \mathbf{H}^\mathcal{T} 
        ( \mathbf{H} \mathbf{G} \bm{\Phi}_{\mathbf{rr}}^{-1} \mathbf{G}^\mathcal{T}  \mathbf{H}^\mathcal{T}  + \rho \mathbf{I} )^{-1} ,
    \label{eq:A_matrix}
\end{align}
where a small regularization factor $\rho$ is included to ensure the numerical stability of the matrix inversion when computing $\mathbf{A}$.

\section{Soft-constrained SSANC in Time Domain} 
\label{sec:soft_constrained_ssanc_in_time_domain}
Since the optimization problem in~\labelcref{eq:ssanc_optimization_problem_both} aims to perfectly preserve the speech component of the reference microphone signal at the inner error microphone, the noise reduction may be limited. By allowing a small amount of speech distortion, i.e., relaxing the distortionless constraint, a larger amount of noise reduction may be achieved, thus potentially improving signal-to-noise ratio and perceived speech quality and intelligibility.

To allow a trade-off between speech distortion and noise reduction, we propose a soft-constrained SSANC, by defining the optimization problem as
\begin{align}
    \Aboxed{
    \min_{\mathbf{w}} \quad \mathcal{E} \left\{e^{2}(n)\right\}  + \beta \| \mathbf{w} \|_2^2 + \mu \| \mathbf{H}(\mathbf{q} + \mathbf{G} \mathbf{w}) -  \bm{\updelta}_{\Delta} \|_2^2 ,
    }
    \label{eq:ssanc_optimization_problem_soft_both}
\end{align}
where $\mu$ is a real-valued non-negative trade-off parameter. 
The solution is found to be
\begin{align}
    \mathbf{w}_\mathrm{soft} =    
        \!-\! \left( \bm{\Phi}_{\mathbf{rr}} \! + \! \mu \mathbf{G}^\mathcal{T} \! \mathbf{H}^\mathcal{T} \! \mathbf{H} \mathbf{G} \right)^{-1} \!\!
        \left[ \bm{\upphi} \!-\! \mu \mathbf{G}^\mathcal{T} \! \mathbf{H}^\mathcal{T} \! ( \bm{\updelta}_{\Delta} \!\!-\! \mathbf{H} \mathbf{q}) \right] \!  .
    \label{eq:w_ssanc_time_soft}
\end{align}

When $\mu = 0$, the optimization problem in~\labelcref{eq:ssanc_optimization_problem_soft_both} becomes equal to~\labelcref{eq:anc_optimization_time_both}, corresponding to the conventional ANC. Noise reduction is maximized in this case, but the desired speech component is also strongly suppressed (see simulations in~\Cref{sec:simulation_results}). Increasing $\mu$ gives more weight to the distortion term in~\labelcref{eq:ssanc_optimization_problem_soft_both}, reducing speech distortion but potentially limiting the amount of noise reduction.

\section{Frequency-Domain Analysis}
\label{sec:SD_SSANC_freq}
In this section, we present a frequency-domain analysis for the conventional ANC, the hard-constrained SSANC, and the proposed soft-constrained SSANC, further highlighting their relationships.

Similarly to~\labelcref{eq:anc_optimization_time_both}, the optimization problem for conventional ANC in the frequency domain is defined as
\begin{align}
    \min_{\mathbf{w}(\omega)} \quad \mathcal{E} \left\{ | E(\omega) |^2 \right\} ,
    \label{eq:anc_cost_func_freq}
\end{align}
where the regularization term is omitted for brevity (i.e., $\beta=0$).
The solution can be found to be
\begin{align}
    \mathbf{w}_\mathrm{ANC}(\omega) 
        &= 
            \frac{ 1 }{ G^*(\omega) }  \left( - \mathbf{q}_{\, \omega}  \right) ,
    \label{eq:anc_w_freq}
\end{align}
where $(\cdot)^*$ denotes the complex conjugate.

Similarly to~\labelcref{eq:ssanc_optimization_problem_both}, the optimization problem for hard-constrained SSANC in the frequency domain is defined as
\begin{subequations}
    \begin{align}
        \min_{\mathbf{w}(\omega)} \quad & \mathcal{E} \left\{ | E(\omega) |^2 \right\}
        \\
        \mathrm{subject \ to} \quad & \left[ \mathbf{q}^\mathcal{H}_{\, \omega} + G(\omega) \mathbf{w}^\mathcal{H}(\omega) \right] \mathbf{h}(\omega) = \exp{(-\mathrm{i} \omega \Delta)} ,
    \end{align}
\end{subequations}
where $\mathbf{h}(\omega) = [H_1(\omega) \dots  H_K(\omega)  H_{K+1}(\omega) ]^\mathcal{T} \! \in \mathbb{C}^{K+1} $ denotes the relative transfer function vector of the desired source, and $\mathrm{i}$ denotes the imaginary unit.
The solution is found to be
\begin{align}
    \mathbf{w}_\mathrm{hard}(\omega) 
        &= 
            \frac{ 1 }{ G^*(\omega) } 
            \left(
            - \mathbf{q}_{\, \omega} 
            +  
            \frac{\bm{\Phi}_{x}^{-1}(\omega) \mathbf{h}(\omega) \exp{(\mathrm{i} \omega \Delta)} }{ \,\, \mathbf{h}^\mathcal{H}(\omega) \bm{\Phi}_{x}^{-1}(\omega) \mathbf{h}(\omega) \,\, }  
            \right) ,
    \label{eq:w_ssanc_freq_hard}
\end{align}
where $\bm{\Phi}_{x}(\omega) = \mathcal{E} \{ \mathbf{x}(\omega) \mathbf{x}^\mathcal{H}(\omega) \} \in \mathbb{C}^{(K+1) \times (K+1)} $ denotes the input covariance matrix, which is assumed to be positive-definite.

Similarly to~\labelcref{eq:ssanc_optimization_problem_soft_both}, the optimization problem for the proposed soft-constrained SSANC in the frequency domain is defined as 
\begin{align}
    \min_{\mathbf{w}(\omega)} \, \mathcal{E} \left\{ | E(\omega) |^2 \right\}
    \!+\! \mu \left|  \left[ \mathbf{q}^\mathcal{H}_{\, \omega} + G(\omega) \mathbf{w}^\mathcal{H}(\omega) \right] \!\mathbf{h}(\omega) \!-\! \exp{(-\mathrm{i} \omega \Delta)} \right|^2  \!.
    \label{eq:ssanc_cost_func_freq_soft}
\end{align}
The solution is found to be
\begin{align}
     \mathbf{w}_{\mathrm{soft}}(\omega) 
        &= 
        \frac{1}{G^*(\omega)} 
        \left( 
        - 
        \mathbf{q}_{\, \omega}  
        +
        \frac{\bm{\Phi}_{x}^{-1}(\omega) \mathbf{h}(\omega)  \exp{(\mathrm{i} \omega \Delta)} }{  \,\, \frac{1}{\mu} + \mathbf{h}^\mathcal{H}\bm{\Phi}_{x}^{-1}(\omega)\mathbf{h}(\omega) \,\,} 
        \right)  .
    \label{eq:w_ssanc_freq_soft}
\end{align}

By comparing~\labelcref{eq:anc_w_freq,eq:w_ssanc_freq_hard,eq:w_ssanc_freq_soft}, it can be observed that
\begin{subequations}
    \begin{align}
        \lim_{\mu \rightarrow 0} \mathbf{w}_{\mathrm{soft}}(\omega) &= \mathbf{w}_{\mathrm{ANC}}(\omega),  
        \\
        \lim_{\mu \rightarrow \infty} \mathbf{w}_{\mathrm{soft}}(\omega) &= \mathbf{w}_{\mathrm{hard}}(\omega),
    \end{align}
    \label{eq:w_soft_limits}
\end{subequations}
implying that in the frequency domain the conventional ANC and the hard-constrained SSANC are limiting cases of the soft-constrained SSANC for $\mu \rightarrow 0$ and $\mu \rightarrow \infty$, respectively.

\section{Simulation Results}
\label{sec:simulation_results}

\subsection{Setup}
\begin{figure}[t]
    \centering
    \captionsetup[subfloat]{captionskip=0pt,farskip=0pt}
    \subfloat[]{\includegraphics[height=0.2\textwidth]{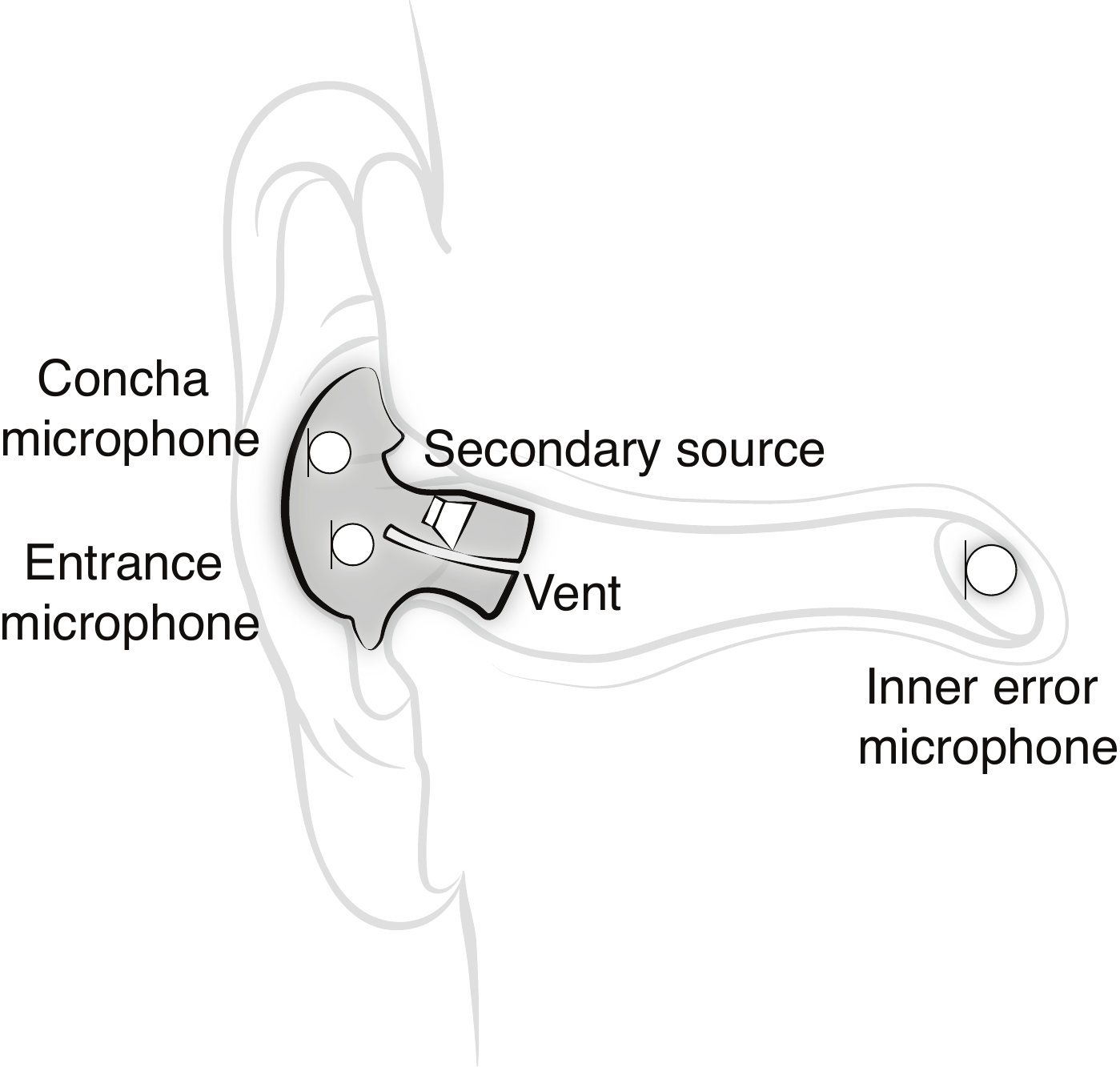} \label{fig:Ear}}
    \subfloat[]{\includegraphics[height=0.2\textwidth]{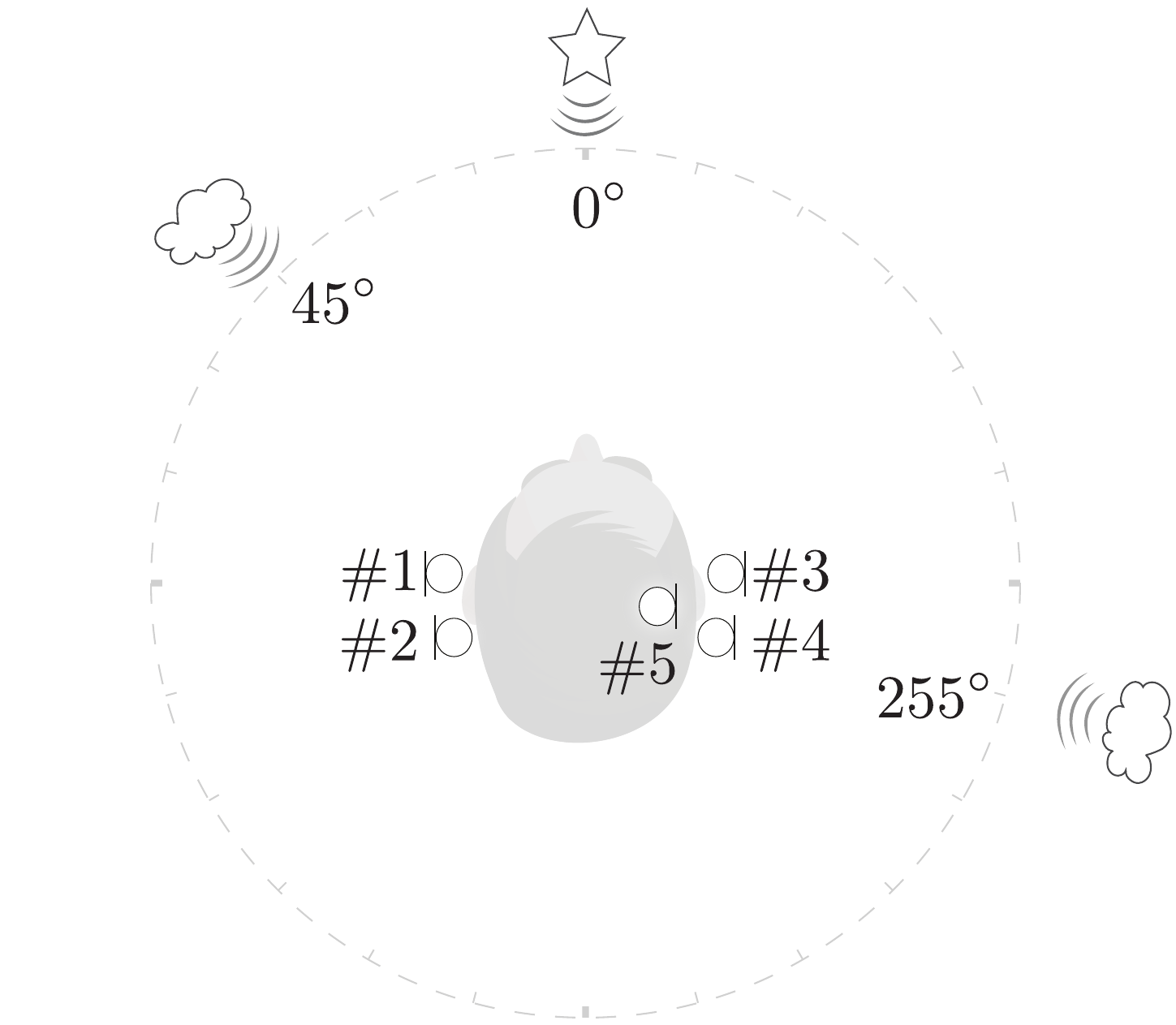} \label{fig:Scenario}}
    \caption{(a) Illustration of the open-fitting hearable. (b) Acoustic scenario with one desired speech source at $0^\circ$ and two babble noise sources at $45^\circ$ and $255^\circ$. }
    \label{fig:Setup}
\end{figure}

For the simulation, we considered a pair of open-fitting hearables~\cite{denk2019one,denk2021hearpiece} inserted into both ears of a GRAS 45BB-12 KEMAR Head \& Torso simulator, as shown in~\cref{fig:Ear}. We used four outer microphones (entrance microphones and concha microphones at the left and right ears, labeled as \#1-- \#4 in \cref{fig:Scenario}), one inner error microphone (located at the right ear, labeled as \#5 in \cref{fig:Scenario}), and the outer receiver at the right ear as the secondary source. The inner error microphone was assumed to be at the eardrum. 

The acoustic scenario is shown in~\cref{fig:Scenario}, where we considered a desired speech source at $0^\circ$ (\say{p361\_005} from the VCTK dataset\cite{veaux2017cstr}) and two noise sources at $45^\circ$ and $255^\circ$ (babble noise from the NOISEX-92 database~\cite{Varga1993assessment}). The desired speech and noise components in all microphone signals were generated by convolving source signals with measured anechoic impulse responses from the database~\cite{denk2021hearpiece}. All signals were 5 seconds in duration and sampled at 16~kHz. The desired speech and noise components were mixed such that the signal-to-noise ratio (SNR) of the leakage at the inner error microphone was set to $-5$~dB, with both noise sources contributing the same energy. As the secondary path estimate we used the measured impulse response between the outer receiver and the inner error microphone at the right ear (from~\cite{denk2021hearpiece}).

All the algorithms were implemented in the time domain. As filter lengths we used $L_w=600$ for the control filter (per channel), $L_g=280$ for the secondary path, and $L_a=22$ and $L_h=262$ for the anti-causal and causal parts of the ReIRs for the desired speech source in~\labelcref{eq:Hk_time}. The delay for the desired speech component in~\labelcref{eq:es_alpha_x_ref_time} was set to $\Delta=32$, corresponding to 2~ms. The ReIRs for the desired speech source were computed using the least-mean-squares adaptive filter (after convergence) from microphone signals generated by convolving  white noise with the measured impulse responses at $0^\circ$, using the entrance microphone \#3 as the reference microphone. Similar to the previous studies~\cite{xiao2023spatial, Xiao2024icassp, Xiao2025fa}, the evaluation focused on the frequency range above 100~Hz. Therefore, a minimum-phase high-pass filter with a cut-off frequency at 100~Hz was applied to the desired signal $x_{\mathrm{ref},s}(n-\Delta)$. For all ANC designs, $\beta = \nicefrac{ \lambda_\mathrm{max} (\mathbf{G}^\mathcal{T} \bm{\Phi}_{{\mathbf{x}} {\mathbf{x}}} \mathbf{G} ) }{ (\num{4e5}) }$, where $\lambda_\mathrm{max}(\cdot)$ denotes the largest eigenvalue. The regularization factor in~\labelcref{eq:A_matrix} was equal to $\rho = \nicefrac{ \lambda_\mathrm{max} (\mathbf{H} \mathbf{G} \bm{\Phi}_{\mathbf{rr}}^{-1} \mathbf{G}^\mathcal{T} \mathbf{H}^\mathcal{T}) }{ (\num{4e5}) }$.

\subsection{Evaluation metrics}
The performance of the considered ANC designs was evaluated in terms of noise reduction, speech distortion, SNR improvement, speech quality and intelligibility.

The noise reduction is defined as the difference between the power of the noise component of the leakage $p_v(n)$ (without control) and the noise component of the inner error microphone signal $e_v(n)$ (with control), i.e.,
\begin{equation}
    \mathrm{NR} \ \mathrm{(dB)}  
    = 
        10\log_{10}   \sum\limits_{n=1}^{N}  p_v^2(n) 
        -
        10\log_{10}   \sum\limits_{n=1}^{N} e_v^2(n) ,
\end{equation}
where $N$ denotes the total signal length.

The intelligibility-weighted spectral distortion is used to assess the amount of speech distortion~\cite{Spriet2004spatially, Doclo2007frequency, Serizel2010integrated}. It is defined as
    \begin{align}
        &\mathrm{SD}_\mathrm{intellig} \, \mathrm{(dB)}
            = 
            \sum\limits_{b=1}^{B} I(\omega_b) \, 10\log_{10} \frac{\mathcal{P}_\epsilon (\omega_b)}{\mathcal{P}_{\mathrm{ref},s} (\omega_b)}  ,
        \label{eq:SD_intellig}
    \end{align}
where the band importance function $I(\omega_b)$ expresses the importance of the $b$-th one-third octave band for intelligibility~\cite{ASA1997}, and $B$ denotes the total number of bands. $\mathcal{P}_\epsilon (\omega_b)$ is the power spectral density of $\epsilon(n)$ in the $b$-th band, where $\epsilon(n) = e_s(n) - x_{\mathrm{ref},s}(n-\Delta)$. $\mathcal{P}_{\mathrm{ref},s} (\omega_b)$ is the power spectral density of $x_{\mathrm{ref},s}(n-\Delta)$ in the $b$-th band.

Similar to~\labelcref{eq:SD_intellig}, the intelligibility-weighted SNR improvement is defined as
\begin{align}
    \Delta \mathrm{SNR}_\mathrm{intellig} \, \mathrm{(dB)} 
        \! = \!
        \sum\limits_{b=1}^{B} I(\omega_b) \, \left[ \mathrm{SNR}^\mathrm{on} \! (\omega_b) - \mathrm{SNR}^\mathrm{off} \! (\omega_b) \right]   ,
\end{align}
where $\mathrm{SNR}^\mathrm{on}(\omega_b)$ and $\mathrm{SNR}^\mathrm{off}(\omega_b)$ denote the SNRs in the $b$-th band of the inner error microphone signal (with control) and the leakage (without control), respectively.

In addition, we considered the narrowband perceptual evaluation of speech quality (PESQ)~\cite{Rix2001PESQ} and the extended short-term objective intelligibility (ESTOI)~\cite{Jensen2016algorithm} metrics using $x_{\mathrm{ref},s}(n-\Delta)$ as the reference signal. We evaluated the PESQ and ESTOI differences between the leakage (without control) and the inner error microphone signal (with control).

\subsection{Results and discussion}
\begin{figure}[t]
    \centering
    \includegraphics[width=0.95\linewidth]{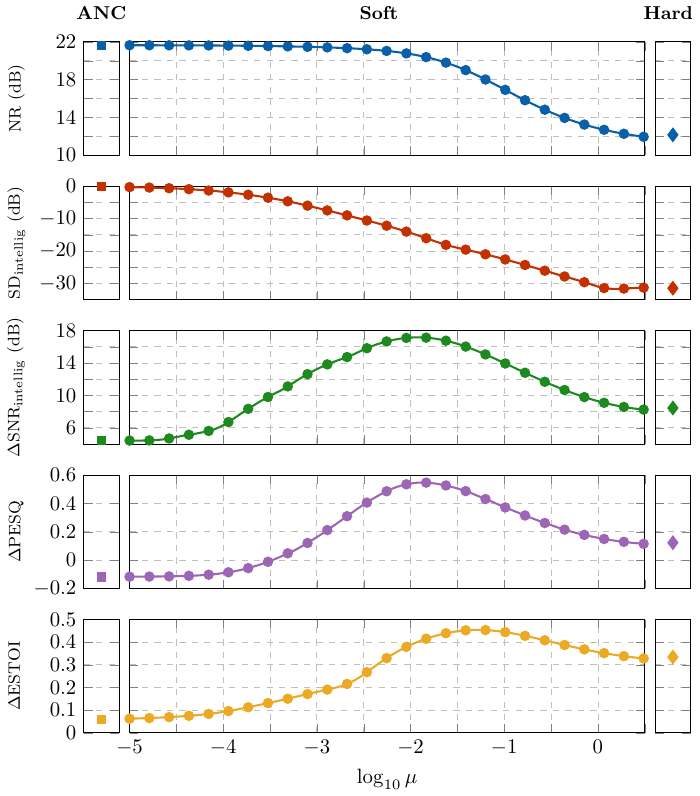}
    \vspace{-8pt}
    \caption{Noise reduction, intelligibility-weighted spectral distortion, intelligibility-weighted SNR improvement, narrowband PESQ improvement, and ESTOI improvement for conventional ANC (left column, $\mu=0$), the hard-constrained SSANC (right column, $\mu \rightarrow \infty$), and the proposed soft-constrained SSANC for different values of the trade-off parameter $\mu$. }
    \label{fig:Result}
\end{figure}

\Cref{fig:Result} compares the performance of the conventional ANC, the hard-constrained SSANC, and the proposed soft-constrained SSANC for different values of the trade-off parameter $\mu$ ($-5 \leq \log_{10}\mu \leq 0.5$). For the conventional ANC, both desired speech and noise components are minimized at the inner error microphone, yielding the maximum noise reduction (21.6~dB) but also significant speech distortion (0~dB). This leads to a poor SNR improvement (4.4~dB), a drop in the PESQ score ($-0.12$), and a minor ESTOI improvement (0.06). In contrast, the hard‑constrained SSANC system almost perfectly preserves the speech component of the reference microphone signal. This results in minimal speech distortion ($-31.5$~dB) while reducing noise by 12.1~dB. An SNR improvement of 8.4~dB, a PESQ improvement of 0.12, and an ESTOI improvement of 0.33 are obtained.

The proposed soft‑constrained SSANC system bridges the gap between both extreme cases. For very small values of $\mu$, it behaves like the conventional ANC system, whereas for large $\mu$ it approaches the hard‑constrained SSANC system, thus confirming the frequency-domain relationships in~\labelcref{eq:w_soft_limits}. More importantly, there is a wide range of parameter values yielding a larger noise reduction than the hard-constrained case without causing excessive speech distortion. For example, setting $\log_{10}\mu = -2$ ($\mu = 0.01$) yielded a noise reduction of 20.7~dB and a speech distortion of $-14.7$~dB, corresponding to an SNR improvement of 17.2~dB, a PESQ improvement of 0.54, and an ESTOI improvement of 0.39, outperforming hard‑constrained SSANC by 8.8~dB in terms of SNR improvement, 0.42 in terms of PESQ improvement, and 0.06 in terms of ESTOI improvement. These results demonstrate that the soft-constrained SSANC not only generalizes existing approaches but also provides a practical method for improving speech quality and intelligibility\footnote{Audio examples can be found at \href{https://uol.de/p113620}{https://uol.de/p113620}.}.

\section{Conclusion}
\label{sec:conclusion}

In this paper, we proposed a soft-constrained SSANC system employing a tunable parameter to trade off between speech distortion and noise reduction. Through a frequency-domain analysis it was shown that the proposed system is equivalent to conventional ANC and hard-constrained SSANC for extreme values of the trade-off parameter, which was confirmed by simulations in the time domain for a pair of open-fitting hearables. The results demonstrated that there exists a wide range of values for the trade-off parameter leading to better SNR improvement and enhanced speech quality and intelligibility than the hard-constrained configuration. Future work includes evaluating the trade-off parameter in different acoustic scenarios.


\clearpage

\bibliographystyle{IEEEtran}
\bibliography{WASPAA2025_paper}

\end{document}